\documentclass[aps,prx,twocolumn,showpacs,superscriptaddress,groupedaddress]{revtex4}
\pdfoutput=1 

\usepackage{amsmath}
\usepackage{amssymb}
\usepackage{docs}%
\usepackage{bm}%
\usepackage[colorlinks=true,linkcolor=blue,citecolor=blue,urlcolor=black]{hyperref}%
\usepackage{graphicx}
\usepackage[utf8]{inputenc}
\usepackage{epstopdf}
\usepackage{mdframed}
\usepackage{xcolor}
\mdfdefinestyle{MyFrame}{%
    linecolor=black,
    outerlinewidth=20pt,
    roundcorner=30pt,
    innertopmargin=\baselineskip,
    innerbottommargin=\baselineskip,
    innerrightmargin=30pt,
    innerleftmargin=30pt,
    backgroundcolor=gray!10!white}
\expandafter\ifx\csname package@font\endcsname\relax\else
 \expandafter\expandafter
 \expandafter\usepackage
 \expandafter\expandafter
 \expandafter{\csname package@font\endcsname}%
\fi
\begin{document}

\title[]{
On the physisorption of water on graphene:
Sub-chemical accuracy from many-body electronic structure methods
} 

\author{Jan Gerit Brandenburg}
\affiliation{Department of Physics and Astronomy, University College London, Gower Street, London WC1E 6BT, United Kingdom}
\affiliation{Thomas Young Centre and London Centre for Nanotechnology, 17-19 Gordon Street, London WC1H 0AH, United Kingdom}
\author{Andrea Zen}
\affiliation{Department of Physics and Astronomy, University College London, Gower Street, London WC1E 6BT, United Kingdom}
\affiliation{Thomas Young Centre and London Centre for Nanotechnology, 17-19 Gordon Street, London WC1H 0AH, United Kingdom}
\author{Martin Fitzner}
\affiliation{Department of Physics and Astronomy, University College London, Gower Street, London WC1E 6BT, United Kingdom}
\affiliation{Thomas Young Centre and London Centre for Nanotechnology, 17-19 Gordon Street, London WC1H 0AH, United Kingdom}
\author{Benjamin Ramberger}
\affiliation{University of Vienna, Faculty of Physics and Center for Computational Materials Sciences, Sensengasse 8/12, 1090 Wien, Austria}
\author{Georg Kresse}
\affiliation{University of Vienna, Faculty of Physics and Center for Computational Materials Sciences, Sensengasse 8/12, 1090 Wien, Austria}
\author{Theodoros Tsatsoulis}
\affiliation{Institute for Theoretical Physics, Vienna University of Technology, Wiedner Hauptstrasse 8-10, 1040 Vienna, Austria}
\affiliation{Max Planck Institute for Solid State Research, Heisenbergstrasse 1, 70569 Stuttgart, Germany}
\author{Andreas Gr\"uneis}
\affiliation{Institute for Theoretical Physics, Vienna University of Technology, Wiedner Hauptstrasse 8-10, 1040 Vienna, Austria}
\affiliation{Max Planck Institute for Solid State Research, Heisenbergstrasse 1, 70569 Stuttgart, Germany}
\author{Angelos Michaelides}
\email{angelos.michaelides@ucl.ac.uk}
\affiliation{Department of Physics and Astronomy, University College London, Gower Street, London WC1E 6BT, United Kingdom}
\affiliation{Thomas Young Centre and London Centre for Nanotechnology, 17-19 Gordon Street, London WC1H 0AH, United Kingdom}
\author{Dario Alf\`{e}}
\email{d.alfe@ucl.ac.uk}
\affiliation{Department of Earth Sciences, University College London, Gower Street, London WC1E 6BT, United Kingdom}
\affiliation{Thomas Young Centre and London Centre for Nanotechnology, 17-19 Gordon Street, London WC1H 0AH, United Kingdom}
\date{\today}

\begin{abstract}
Molecular adsorption on surfaces plays a central role in catalysis, corrosion, desalination, and many other processes of relevance to industry and the natural world. 
Few adsorption systems are more ubiquitous or of more widespread importance than those involving water and carbon, and for a molecular level understanding of such interfaces water monomer adsorption on graphene is a fundamental and representative system. 
This system is particularly interesting as it calls for an accurate treatment of electron correlation effects, as well as posing a practical challenge to experiments.
Here, we employ many-body electronic structure methodologies that can be rigorously converged and thus provide faithful references for the molecule-surface interaction. 
In particular, we use diffusion Monte-Carlo (DMC), coupled cluster (CCSD(T)), as well as the random phase approximation (RPA) to calculate the strength of the interaction between water and an extended graphene surface. 
We establish excellent, sub-chemical, agreement between the complementary high-level methodologies, and an adsorption energy estimate in the most stable configuration of approximately -100\,meV is obtained. 
We also find that the adsorption energy is rather insensitive to the orientation of the water molecule on the surface, despite different binding motifs involving qualitatively different interfacial charge reorganisation.
In producing the first demonstrably accurate adsorption energies for water on graphene this work also resolves discrepancies amongst previously reported values for this widely studied system.
It also paves the way for more accurate and reliable studies of liquid water at carbon interfaces with cheaper computational methods, such as density functional theory and classical potentials. %
\end{abstract}

\maketitle 

\bigskip
Subject area: Chemical Physics


\section{Introduction}
\label{sec:intro}

The adsorption and diffusion of molecules on surfaces is central to countless industrial applications, including 
catalysis, gas storage, desalination, and more. 
Of all the many and varied molecular adsorption systems, few, if any, are of greater importance than those involving water and carbon. 
Such interfaces are, for example, at the very heart of water purification and desalination membranes. 
In addition, water-carbon interfaces are incredibly interesting scientifically in that they can exhibit unique and fascinating behavior \cite{Fumagalli2018_dielectric, Abraham2017, Joshi2014_MolSieving, Nair2012, AlgaraSiller:2015_2Dice, water_cnt_secchi, md_w@capillaries, Bocquet-NatRevChem-2017, bocquet-naturecomm-2018}. 
For example, water can flow in an essentially frictionless manner across the surfaces of sp$^2$-bonded carbon materials (both carbon nanotubes and graphene) \cite{Hummer:2001gd, Strogatz:2005kj, Holt:2006kr, water_cnt_secchi}.
However, seemingly similar materials such as nanotubes made from hexagonal boron nitride do not exhibit such behaviour~\cite{water_cnt_secchi, md_w@capillaries, Tocci2014_friction, Michaelides:2016_Nature_news, Esfandiar:2017_transport}. 
\begin{figure}[t]
\centering
\includegraphics[width=0.45\textwidth]{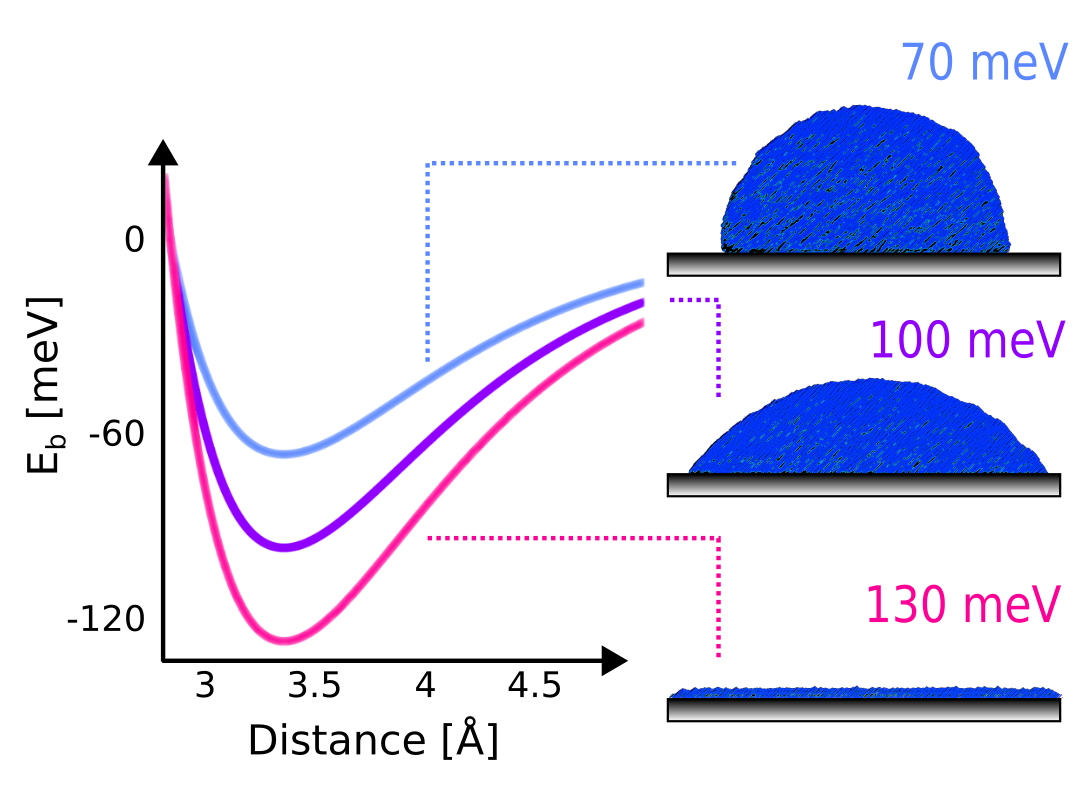}
\caption{\label{fig:angle} 
Water droplet modeled by a coarse-gained water model (mW) and Morse-type water-wall potential with varying adsorption strengths (see appendix~\ref{app:angle}). This figure illustrates that modest changes in the adsorption energy lead to drastic changes in the wetting properties of the surface.
}
\end{figure}
Apart from water purification, wetting of surfaces is of general interest and graphene can be seen as representative van der Waals (vdW) material. 
Many of these exciting experiments showing exceptional properties of water on graphitic surfaces and the outstanding applications lack a full molecular level understanding of the processes and the mechanisms involved. 
In order to gain further insight, it is necessary to complement experimental investigations with theoretical approaches.
The most fundamental property of any adsorption system is the question of the adsorption bond itself, what the strength of the interaction is, the orientation of the molecule, and the physical nature of the interaction. 
For water-carbon interfaces at their simplest level, this comes down to understanding how a single water molecule bonds to a single sheet of graphene.
Accurately establishing the strength of this interaction is important because it directly impacts upon the behavior of water at graphitic surfaces. 
For instance, as illustrated in Fig.~\ref{fig:angle}, molecular dynamics simulations (see appendix~\ref{app:angle}) reveal that altering the strength of the water-graphene interaction by as little as 60\,meV transforms the graphene surface from hydrophobic to hydrophilic. 

Experimentally, water molecules readily form clusters on surfaces and graphene is no exception~\cite{water-interface}.
This has so far prohibited single molecule adsorption measurements of water on graphene. 
On the other hand, the subtle balance of the contributing intermolecular interactions and the very high accuracy required makes the theoretical description by electronic structure methods challenging. 
Indeed, adsorption energy estimates have to be more accurate than so called ``chemical accuracy'', which is defined as 1\,kcal/mol or 43\,meV, in order to be useful in this system.
This is often beyond the accuracy delivered by density functional theory (DFT); the most widely used electronic structure method for understanding molecules at surfaces.
Indeed, previous work has shown that DFT can provide any value for the water adsorption strength on graphene between 0 and -160\,meV depending on the exchange-correlation functional and vdW-correction~\cite{water@graphene_dmc,water@graphene_dft1, water@graphene_dft2, water@graphene_dft3}.
\begin{table}[t]
\caption{
Equilibrium adsorption energies $E_{\text{ad}}$ in meV of a single water monomer (in the so-called ``2-leg'' configuration; see Fig. 2) on graphene as estimated from various electronic structure methods. A brief comment is also made about each adsorption energy estimate.
}\label{tab:literature}
\begin{ruledtabular}
\begin{tabular}{l l p{4.5cm}}
 $E_{\text{ad}}$/meV & Method & Comment\\
\cline{1-3}\\[-0.2cm]
 -130 & DFT/CC \cite{water@graphene_dft/cc} & Corrects DFT based on differences on small clusters\\
 -130 & DFT-SAPT \cite{water@graphene_sapt2} & Extrapolation from cluster\\
 -70 $\pm$ 10\footnotemark[2] & DMC \cite{water@graphene_dmc} & Periodic system, 
large stochastic error, finite-size effects \\
 -135 & i-CCSD(T) \cite{water@graphene_paulus} & Incremental expansion, correlation from
cluster, small basis set \\
 -87  & p-CCSD(T) \footnotemark[1] & Periodic system,
finite-size corrected \\
 -99 $\pm$ 6\footnotemark[2] & DMC \footnotemark[1] & Periodic system,
finite-size
corrected \\
\end{tabular}
\footnotetext[1]{This work.}
\footnotetext[2]{Error due to DMC stochastic uncertainty.}
\end{ruledtabular}
\end{table}%

The variation in adsorption energies obtained from DFT calls for many-body methodologies that can be rigorously converged.
The application of such methods to extended (periodic) surfaces, however, does not come without significant challenges~\cite{advancesQC_editorial}. %
Table~\ref{tab:literature} summarizes attempts to provide  benchmark quality binding energies for water on graphene with state-of-the-art electronic structure methods ~\cite{water@graphene_dft/cc,water@graphene_sapt2,water@graphene_dmc,water@graphene_paulus,water@graphene_mp2}.
It can be seen that the previous estimates range from about 70 to about 130 meV. 
Whilst this is narrower than the range obtained from DFT, the deviations are clearly too high to make faithful predictions on the water-graphene interactions.
Although these previous attempts have involved great care and considerable effort, they all have possible  weaknesses and potential shortcomings. 
Here, in light of a new estimate of the adsorption energy of water on graphene, we will critically examine previous estimates and in so-doing resolve the discrepancies between previous reports.
Our new study relies on impressive progress with state-of-the-art electronic structure methods and their implementation in scaleable software suits in the past few years. 
These developments together with an increased capacity of available computational resources  makes the accurate determination of binding energies on extended surfaces feasible. 
It is thus timely to analyze the physisorption of water on graphene again and in so-doing we are able to demonstrate that different many-body electronic structure methods indeed agree within sub-chemical accuracy.

In the remainder of this manuscript we briefly discuss the water monomer binding motifs considered and summarize the main methodologies employed (Section~\ref{sec:methods}). 
In Section~\ref{sec:results}, we summarize and discuss our best estimates
for the adsorption energies (\ref{sec:adsorption_ref}) of water on a periodic free-standing graphene sheet and on small sp$^2$-bonded molecular analogues of graphene (specifically benzene and coronene).
Following this we discuss the discrepancies between previous literature values (\ref{sec:issues}), the implications of our revised adsorption energies including an assessment of the performance of certain DFT functionals (\ref{sec:implications}), and analyze the physical nature of the adsorption bond (\ref{sec:analyze}).
Conclusions and a future perspective are given in section~\ref{sec:conclusion}.

\section{Methods}
\label{sec:methods}
Various computational approaches have been used in the current study. 
Here, mainly for orientation purposes, we briefly comment on the structures examined, the quantities computed, and the main methods used.  Full details of all methods are provided in Appendix~\ref{sec:appendix-compdetail}. 

\begin{figure}[th]
\centering
\includegraphics[width=0.48\textwidth]{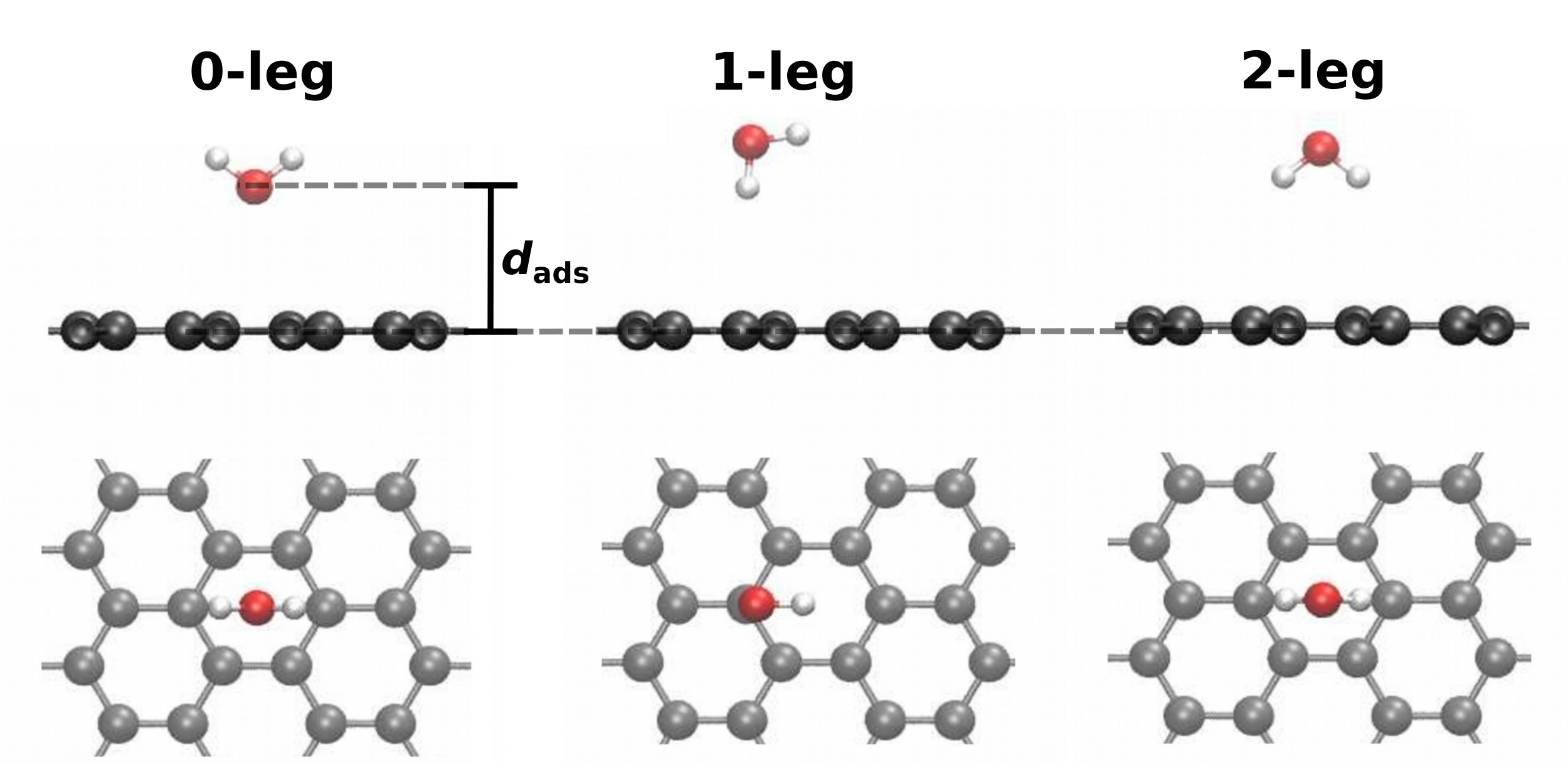}
\caption{\label{fig:dmc_dario} 
Water adsorption structures considered. We show the 0-leg, 1-leg, and 2-leg motif each from the side (top) and from above (bottom). 
The distance $d$ is defined by the  distance of the oxygen atom from the graphene plane.
A periodic 5$\times$5 graphene cell is used in most calculations, while for clarity only a small part of the simulation cell is shown.
All equilibrium geometries are provided as supporting information.
}
\end{figure}

Water monomer adsorption was considered in three different motifs, dubbed 0-leg, 1-leg, and 2-leg as defined in Fig.~\ref{fig:dmc_dario}. 
These are the most widely discussed adsorption structures in previous studies~\cite{water@graphene_dft/cc,water@graphene_sapt2,water@graphene_dmc,water@graphene_paulus,water@graphene_mp2}. 
The adsorption energy $E_{\text{ad}}$ is 
the minimum, obtained at the equilibrium distance $d_\text{ad}$, of the binding curve: 
\begin{align}
    E_{\text{b}}(d)=E_{\text{W+G}}(d)-E_{\text{W+G}}(d_\text{far}) \,,
    \label{eq:ead}
\end{align}
where $E_{\text{W+G}}(d)$  is the energy of the system with the water at a distance $d$ from the graphene sheet, and $E_{\text{W+G}}(d_\text{far})$ is the energy of the non-interaction system, with water far away at a distance $d_\text{far}$ from graphene~\footnote{Ideally, the energy for non-interacting systems $E_{\text{W+G}}(d_\text{far})$ should be evaluated at a distance $d_\text{far} \to \infty$. Computationally, there are two alternative ways to calculate $E_{\text{W+G}}(d_\text{far})$: {\em i}. consider a finite value of $d_\text{far}$ that makes the residual interaction energy negligible with respect to the needed accuracy; {\em ii}. evaluate as the summation of the energy of the water and graphene fragments. The two methods are in principle equivalent for size-consistent electronic structure approaches, however in periodic systems (so, in presence of finite size errors) the former method typically benefits from a better error cancellation and allows the use of smaller simulation cells. Details are discussed in Appendix~\ref{sec:appendix-compdetail}.}.
Note that what is computed here is the static electronic energy without inclusion of thermal or nuclear quantum effects.

One of the key techniques used in this study is diffusion Monte-Carlo (DMC)~\cite{foulkes01}. %
DMC can be readily applied to periodic systems and in the past few years the computational efficiency and accuracy of the technique has improved significantly. 
In particular, a recent algorithmic development has reduced computational effort by up to two orders of magnitude~\cite{sizeconsDMC}. 
Subsequently it was shown that the new DMC algorithm together with an effective estimation of finite-size errors yields chemically accurate lattice energies for a range of molecular crystals (including ice and delocalized $\pi$-systems) with modest computational cost~\cite{molcryst_dmc}.
In the current study we performed DMC studies with the CASINO code~\cite{casino} for water on benzene, coronene, and graphene in periodic unit cells.
For adsorption on graphene a large 5$\times$5 unit cell
was employed.

The second many-body approach used in this study is coupled cluster, for both water adsorption on benzene and coronone and on periodic graphene.  
Specifically, we used linear scaling domain based pair natural orbital coupled cluster including singles, doubles, and perturbative triples [denoted here as the L-CCSD(T) method~\cite{dlpnoccsdt}], as implemented in the ORCA program package~\cite{orca} for the finite sized cluster models.
Periodic  coupled cluster including singles, doubles, and perturbative triples [denoted here as the p-CCSD(T)] was used for adsorption on graphene~\cite{solid_exact}. 
For these calculations a periodic 4$\times$4 unit cell was employed and 
the coupled cluster code CC4S interfaced to the VASP code~\cite{vasp2,vasp3} was used~\cite{J-factorization, pbc-ccsdt-Nlimit}.

The third many-body approach employed is the random phase approximation (RPA)~\cite{RPA:1958, RPA:Kresse2008, RPA_Kresse:PRL2009,rpa_furche,rpa_cubic_hutter}, which computes the correlation energy based on the electron density response function.
RPA is computationally more affordable than CCSD(T) and has recently shown good results, in particular if singles corrections are introduced~\cite{RPA_Schimka:NatMatt2010, Ren:2011ht, Jiri:2016, molcryst_dmc}. 
However, it includes fewer excitation types than CCSD(T) and thus one has to carefully test it's accuracy.
Here we used RPA based on PBE orbitals, i.e.\ 
the exact exchange energy EXX@PBE combined with the correlation RPA@PBE. 
In addition, the contribution from GW single excitations (GWSEs) was computed based on the work of Klime\v{s} et al.~\cite{vasp_gwse}.

Note that the calculations with the various methods used the same adsorption structures (generated from DFT optimizations), and as reference the isolated fragments with fixed (unrelaxed) geometries are taken (see~\ref{subsec:system}).
Additional details of the computational settings used with the various methods are provided in Appendix~\ref{sec:appendix-compdetail}. 


\section{Results}
\label{sec:results}

\subsection{High-level adsorption energies}
\label{sec:adsorption_ref}
\begin{figure}[hbt]
\centering
\includegraphics[width=0.5\textwidth]{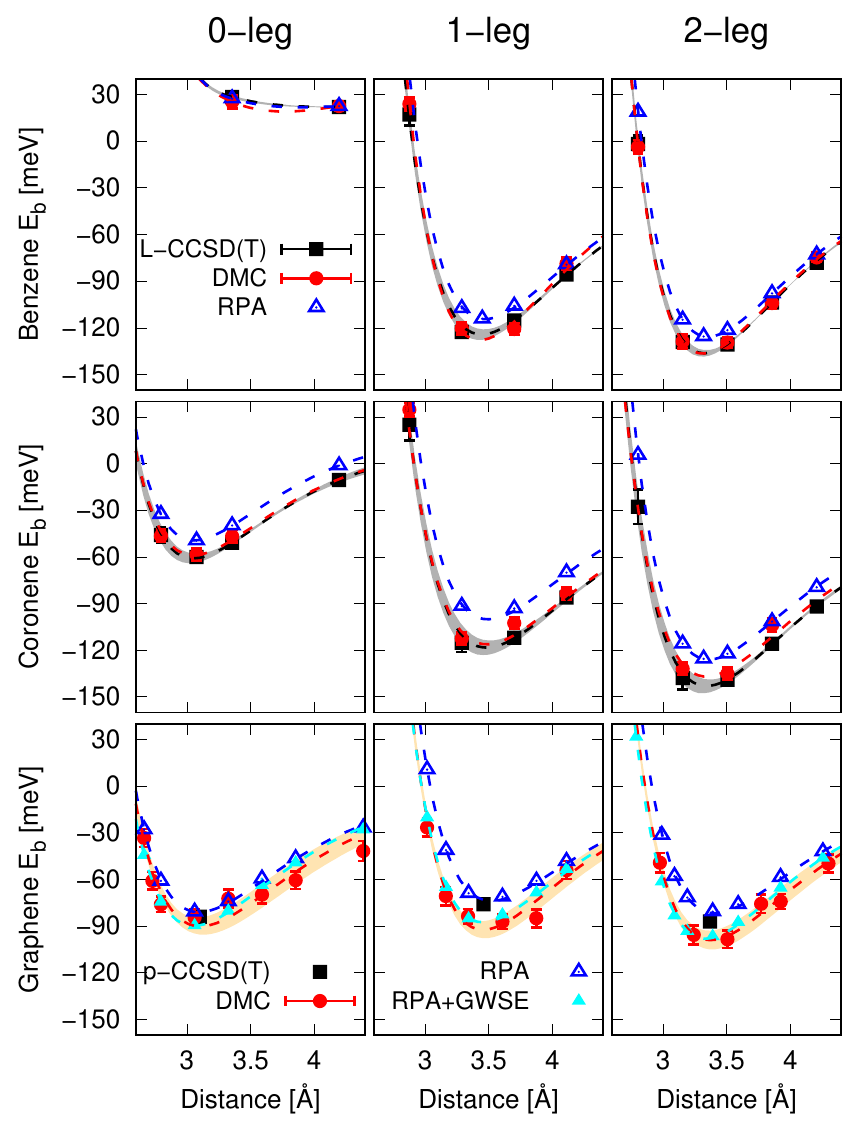}
\caption{\label{fig:pes} 
Binding energy curves of water on
benzene (top), coronene (middle), and graphene (bottom) in the 0-leg, 1-leg, and 2-leg motifs
computed with the many-body electronic structure methods L-CCSD(T), DMC, and RPA.
Error bars in DMC correspond to the stochastic error (1 standard deviation), in L-CCSD(T) to the basis set uncertainty
(see Appendix~\ref{sec:appendix-compdetail}).
Dashed lines are fits via a Morse potential (see Appendix \ref{sec:appendix-morses}).
}
\end{figure}
Interaction energy curves of water adsorbed in three different configurations (0-leg, 1-leg, 2-leg, see Fig.~\ref{fig:dmc_dario}) on benzene, coronene, and graphene have been computed with a range of  many-body methods. 
Here, DMC as well as CCSD(T) are considered benchmark quality methods, while RPA is tested as a cheaper alternative. %
Due to the smaller unit cell used in p-CCSD(T), we expect that DMC provides the best reference interaction energies for water adsorbed on graphene. 
Combined interaction energy curves for the considered systems are shown in Fig~\ref{fig:pes},
and interpolated minima are given in Table~\ref{tab:energies}.

We begin by noting that the 0-leg configuration on benzene is purely repulsive (i.e. unbound), while it gets increasingly attractive on coronene and graphene. 
This behavior will be rationalized in terms of the charge density distribution in section~\ref{sec:analyze}.
This trend is consistent with all methodologies and DMC and L-CCSD(T) agree within 2\,meV for this binding motif. 
In contrast, both the 1-leg and 2-leg structures bind on benzene, with the 2-leg adsorption 5-11\,meV stronger. 
This binding energy difference increases on coronene to 15-22\,meV. 
This is due to a decreased binding energy of the 1-leg motif (from benzene to coronene), while the 2-leg binding is identical from DMC and RPA and even slightly stronger from L-CCSD(T).
In stark contrast to the behavior observed on the small molecules, on periodic graphene the 0-leg, 1-leg, and 2-leg structures have very similar DMC binding energies of $-90\pm6$, $-92\pm6$, and $-99\pm6$\,meV, respectively. 
Interestingly, this includes the 0-leg configuration which on benzene was purely repulsive. 
The contrast between benzene and graphene for the 0-leg motif is quite remarkable and will be commented on in more detail in section~\ref{sec:analyze}.

The benchmark quality methods DMC and L-CCSD(T) agree with each other on the molecular clusters with a maximum error of 12\,meV on the single point evaluations and 6\,meV for the interpolated binding energies.
Similarly, the DMC and CCSD(T) equilibrium binding energies on graphene have only small deviations between 6 and 16\,meV.
As the two electronic structure methods have quite distinct foundations, this gives us confidence in the high accuracy of the reported binding energies.
RPA consistently underbinds all structures by about 9-18 meV, this underestimation is effectively reduced by the singles corrections. 
The GW based density corrections also change the relative binding of the three motifs, making the 2-leg 8~meV more stable than the 0-leg (in agreement with DMC and CCSD(T)). 
The observed trend of underestimated RPA binding energies and highly accurate RPA+GWSE energies is consistent with previous findings on molecular adsorption~\cite{water_at_hbn} and molecular crystal lattice energies~\cite{Jiri:2016}.
From the interpolated potential energy curves, we additionally extract the equilibrium adsorption distance and energy (see Table~\ref{tab:energies}). The distances from all many-body methods are in good agreement with each other with a maximum difference of 0.06\,\AA. 
Our best DMC estimate for the water adsorption energy on graphene in the lowest energy 2-leg configuration
($-99$\,$\pm$6\,meV)
is in between previously reported binding energies. 
The p-CCSD(T) adsorption energy is in very good agreement, though the binding is at $-87$\,meV slightly lower.  
This might be due to finite-size or coverage effects (originating from the smaller $4\times 4$ cell employed) and some remaining sensitivity to the basis set. We note that the remaining finite coverage effects have only been corrected for on the level of Hartree-Fock (HF) theory.
\begin{table*}[hbt]
\caption{
Equilibrium binding distances $d_\text{ad}$ and energies $E_{\text{ad}}$ of single water monomer on benzene, coronene, and graphene
from DMC, L-CCSD(T), p-CCSD(T), RPA, and RPA+GWSE. 
Distances are given in \AA\ and energies in meV.
}\label{tab:energies}
\begin{ruledtabular}
\begin{tabular}{l  rr r rr r rr r rr}
	&	\multicolumn{2}{c}{DMC\footnotemark[1]}			&	\phantom{p}	&	\multicolumn{2}{c}{L-CCSD(T)\footnotemark[1]}			&	\phantom{p}	&	\multicolumn{2}{c}{RPA\footnotemark[1]}			\\						
\phantom{{\bf benzene}}	&	$d_\text{ad}$/\AA 	&	$E_{\text{ad}}$/meV  	&		&	$d_\text{ad}$/\AA 	&	$E_{\text{ad}}$/meV  	&		&	$d_\text{ad}$/\AA 	&	$E_{\text{ad}}$/meV  	\\						
		\cline{2-3}						\cline{5-6}						\cline{8-9}									
\multicolumn{9}{l}{\bf benzene}\\																							
0-leg	&	\multicolumn{2}{c}{not binding}			&		&	\multicolumn{2}{c}{not binding}			&		&	\multicolumn{2}{c}{not binding}			\\						
1-leg	&	 3.43(2)	&	 -128(5)	&		&	 3.45(2)	&	 -124(3)	&		&	 3.47(1)	&	 -114(3)	\\						
2-leg	&	 3.31(1)	&	 -136(5)	&		&	 3.32(1)	&	 -136(2)	&		&	 3.35(1)	&	 -126(1)	\\						
\multicolumn{9}{l}{\bf coronene}\\																							
0-leg	&	 ---\footnotemark[2] 	&	   -59(5) 	&		&	 3.05(3) 	&	   -61(3) 	&		&	 3.06(1) 	&	 -49(2)  	\\						
1-leg	&	 ---\footnotemark[2] 	&	 -116(4) 	&		&	 3.48(3) 	&	 -118(5)	&		&	 3.49(1) 	&	 -100(4)  	\\						
2-leg	&	 ---\footnotemark[2] 	&	 -137(4) 	&		&	 3.34(3) 	&	 -143(4) 	&		&	 3.36(1) 	&	 -126(2) 	\\[0.2cm]						
\hline \hline															

	&	\multicolumn{2}{c}{DMC\footnotemark[1]}			&	\phantom{p}	&	\multicolumn{2}{c}{p-CCSD(T)}			&	\phantom{p}	&	\multicolumn{2}{c}{RPA\footnotemark[1]}			&	\phantom{p}	&	\multicolumn{2}{c}{RPA+GWSE\footnotemark[1]}			\\
\phantom{{\bf benzene}}	&	$d_\text{ad}$/\AA 	&	$E_{\text{ad}}$/meV  	&		&	$d_\text{ad}$/\AA 	&	$E_{\text{ad}}$/meV  	&		&	$d_\text{ad}$/\AA 	&	$E_{\text{ad}}$/meV  	&		&	$d_\text{ad}$/\AA 	&	$E_{\text{ad}}$/meV  	\\
		\cline{2-3}						\cline{5-6}						\cline{8-9}						\cline{11-12}
\multicolumn{9}{l}{\bf graphene}\\	
0-leg	&	3.10(3)	&	 -90(6)	&		&	 ---\footnotemark[3] 	&	-84	&		&	 3.09(1)	&	 -81(2)	&		&	 3.05(1)	&	 -90(2)	\\
1-leg	&	3.46(3)	&	 -92(6)  	&		&	 ---\footnotemark[3] 	&	-76	&		&	 3.52(1)	&	 -74(1)	&		&	 3.45(1)	&	 -87(1)	\\
2-leg	&	3.37(2)	&	 -99(6) 	&		&	 ---\footnotemark[3] 	&	-87	&		&	 3.41(1)	&	 -82(1)	&		&	 3.33(1)	&	 -98(1)	\\[-0.3cm]
\footnotetext[1]{Potentials curves (see Fig.~\ref{fig:pes}) are interpolated with a Morse potential, yielding $E_{\text{ad}}$, $d_\text{ad}$, and the corresponding errors that is a combined stochastic and fitting error for DMC (reported in parenthesis relative to the last significant digit, see Appendix~\ref{sec:appendix-morses}).}
\footnotetext[2]{DMC value for $d_\text{ad}$ is not estimated, as the calculated DMC points for water on coronene are not enough for a four parameter fit. We assume that $d_\text{ad}$ is the same as L-CCSD(T) and we estimate only $E_{\text{ad}}$. }
\footnotetext[3]{p-CCSD(T) value calculated only in a single point, at $d_\text{ad}$ 3.10~\AA{} for the 0-leg, 3.46~\AA{} for the 1-leg, 3.37~\AA{} for the 2-leg. }
\end{tabular}
\end{ruledtabular}
\end{table*}%

\subsection{Comparison to literature values}
\label{sec:issues}
As the revised binding energy differs from previously reported values (see Table~\ref{tab:literature}), we carefully investigated the previously used numerical settings and assumptions.
First, the difference with the earlier DMC value of  $-70 \pm 10$~meV~\cite{water@graphene_dmc}, can be explained by the larger statistical errors (smaller precision) and remaining finite-size effects of the older study. 
On both points, the present study has been improved substantially. Taking this into account, both studies agree within their 95\% confidence interval.

The incremental CCSD(T) based adsorption energies are 30 to 35\,meV more strongly bound than the DMC values reported here~\cite{water@graphene_paulus}.
The higher uncertainty of the previous study compared to the DMC and p-CCSD(T) values of this work can be attributed to the basis set employed in the earlier study. 
Specifically, a mixed double-$\zeta$/triple-$\zeta$ basis set expansion was used and as shown in Appendix~\ref{subsec:cc} 
for the water-benzene interaction this does not yield fully converged adsorption energies.  
Although these basis set tests have been performed with benzene as the substrate and some basis set errors at the HF and correlation level partially cancel, the extent of error cancellation cannot be predicted and we expect a significantly increased accuracy from our p-CCSD(T) calculations.

The other studies listed in Table~\ref{tab:literature} use finite-sized clusters to approach the periodic system.
Given the variation in adsorption energies and different nature of the adsorption bond (see section~\ref{sec:analyze}) upon going from benzene, to coronene, to graphene any estimate of the adsorption energy on graphene based on extrapolations from clusters must be done with extreme care. 
\begin{figure}[bht]
\centering
\includegraphics[width=0.49\textwidth]{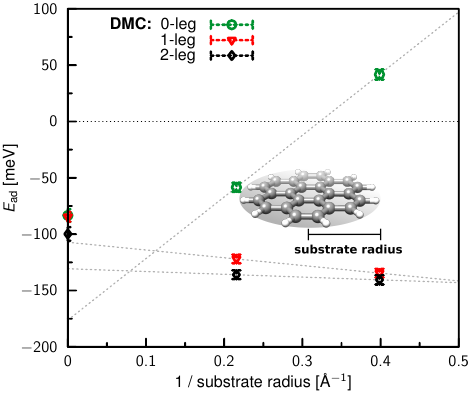}
\caption{\label{fig:convergence} 
Water adsorption energies at fixed equilibrium distances on graphene, coronene, and benzene plotted as a function of  (inverse) substrate size. 
Following the size definition used in Ref.~\cite{substrate-radius} benzene is \ 0.40\,\AA$^{-1}$\ , coronene is 0.22\,\AA$^{-1}$\ , and graphene is 0\,\AA$^{-1}$\.
Dashed lines are fitted using benzene and coronene data.
}
\end{figure}
Whilst slightly different from earlier studies~\cite{water@graphene_dft/cc,water@graphene_sapt2}, we illustrate the challenge in Fig.~\ref{fig:convergence} where the adsorption energy of the three binding motifs is plotted as a function of cluster size~\cite{substrate-radius}.
In particular, from Fig.~\ref{fig:convergence} it can be seen that whilst there is a monotonic convergence towards the adsorption energy on graphene for all adsorption motifs, the extrapolation using benzene and coronene data gives a reasonable result only for the 1-leg structure. 
The extrapolation for both the 0-leg and 2-leg geometries  substantially overestimates the adsorption energy and we see a remarkable sensitivity towards the water orientation. 
Although not reported in the figure this trend is even slightly stronger using the L-CCSD(T) data, where the 2-leg adsorption energy increases from benzene to coronene, i.e.\ a qualitatively different convergence trend towards graphene~\cite{dmc_error}. 
The pronounced contrast in the adsorption energies of the different motifs on small substrates compared to graphene is quite striking.
Overall this suggests for such a delicate system as water on graphene adsorption energies obtained directly on periodic models are likely to be more reliable than those obtained with cluster models. 
\subsection{Implications of revised adsorption energies}
\label{sec:implications}

Using our reliable benchmark interaction energies, we can now test the capability of standard density functional approximations (DFAs) and a few simplified electronic structure methods for water adsorption on graphene. 
\begin{figure}[thb]
\centering
\includegraphics[width=0.49\textwidth]{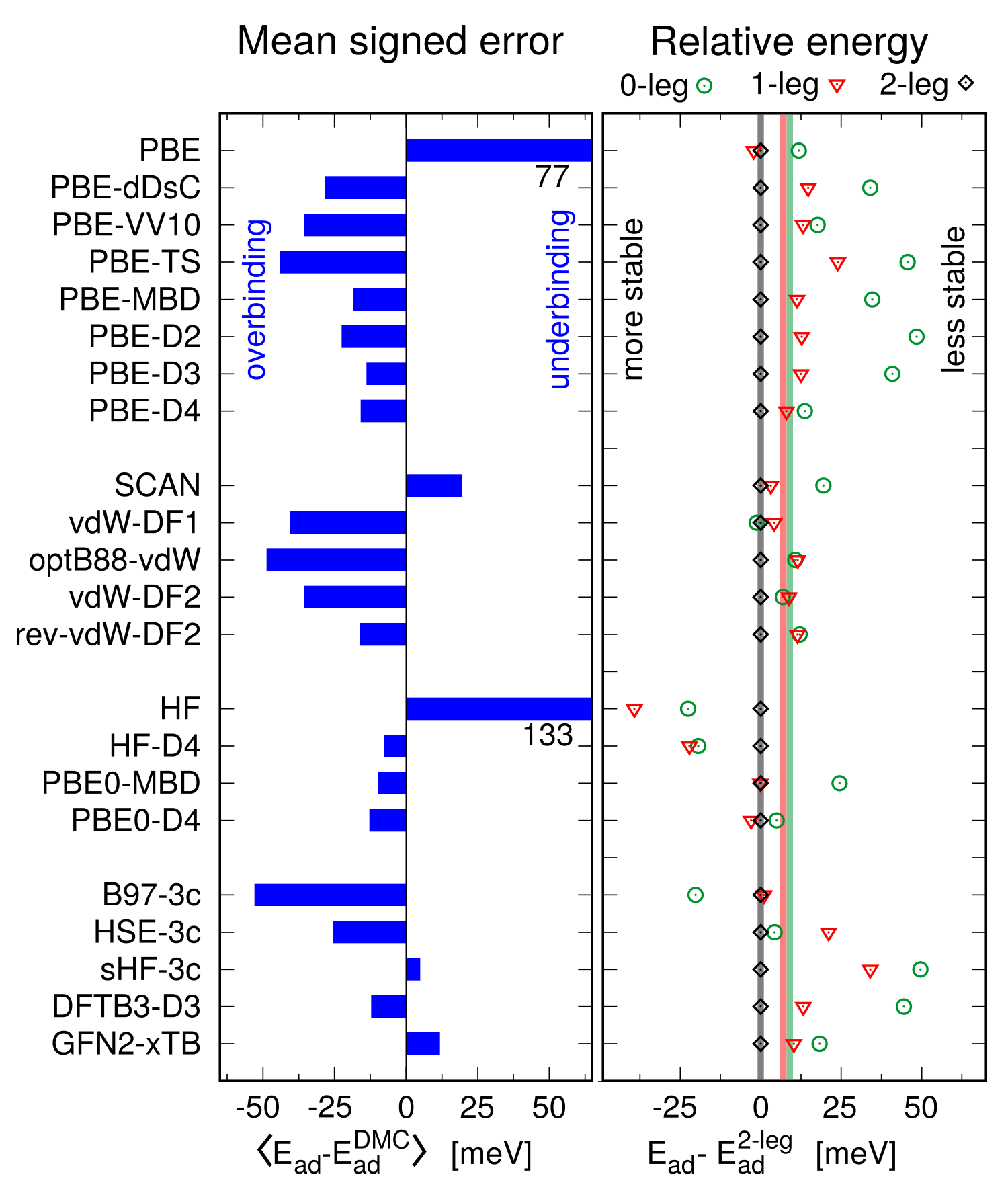}
\caption{\label{fig:bench} 
Mean signed errors of water adsorption energies on graphene computed by various DFAs compared to the DMC references are shown (left panel) as well as the relative stability of the three adsorption motifs (right panel, vertical bars correspond to the DMC references). All energies are computed at fixed DMC equilibrium adsorption distances, reported in Table~\ref{tab:energies}. References to individual methods are given in Appendix~\ref{sec:appendix-compdetail}.
}
\end{figure}
The performance of the various DFAs is summarized in Fig.~\ref{fig:bench}. 
We compare to the DMC references and report the mean signed errors (MEs) as well as the relative adsorption energy of the different motifs. While relative energies can sometimes be easier to compute, the different contributions to the intermolecular interactions make this tougher than getting the average interaction correct.
The plain PBE functional~\cite{pbe} leads to only a rather weak binding for all motifs, clearly long-range vdW interactions are missing. 
Adding a range of appropriate corrections~\cite{vdw_perspective,disp_chemrev,vdwdf_review,hoja_review} clearly improves the binding. However, the error spread is still significant and the older pairwise schemes (PBE-D2~\cite{dftd2}, PBE-TS~\cite{ts}, PBE-dDsC~\cite{ddsc}, and PBE-VV10~\cite{vv10}) cannot be recommended. 
Recent developments with vdW corrections pay off and we can see a clear improvement of PBE-MBD~\cite{ts-mbd} and PBE-D4~\cite{dftd4} over their predecessors.  
The many-body vdW contributions decrease the binding yielding a better agreement with  the DMC references. Most of this effect is already covered at the Axilrod-Teller-Muto type three-body level~\cite{atm1,atm2} as included in the D3 method~\cite{dftd3}. At the PBE-vdW level, only D4 and VV10 are able to reproduce the relative stability to good accuracy within the DMC stochastic error.

The SCAN functional~\cite{scan} is a modern metaGGA that can cover part of the medium-range correlation and has been shown to yield good structures and energies for diversely bonded systems~\cite{scan_natchem}. However, it systematically underbinds all structures, which can partially be compensated by correction schemes~\cite{scand3,scanvv10}.  Still, due to it's intrinsic coverage of some vdW forces, it is non-trivial to combine SCAN with correction schemes~\cite{dfa_disp_balance}. Nonlocal vdW functionals of the first generation (vdW-DF1~\cite{vdwdf}, optB88-vdW~\cite{optpbevdw}) systematically overbind all binding motifs, as seen before for water inside a carbon nanotube~\cite{yasmine_water_cnt}.
This is improved in the second generation and vdW-DF2~\cite{vdwdf2} and rev-vdW-DF2~\cite{revvdwdf2} both give reasonably accurate results. 
We also note that the relative stabilities of the different binding motifs are much better described using the SCAN functional than for PBE.
In order to improve the exchange interactions, fractions of one-determinantal (Fock) exchange are often included in DFAs. %
Indeed, PBE0~\cite{pbe0} combined with the most successful vdW corrections (D4 and MBD)  improve slightly over its GGA pendant.
The simplified DFAs (sHF-3c~\cite{hf3c,shf3c}, HSE-3c~\cite{hse3c}, B97-3c~\cite{b973c})  give mixed results, overall their accuracy is similar to the average dispersion corrected DFA. As they have been designed for increased computational speed (speedup of up to 2 orders of magnitude) they might still be useful for screening applications~\cite{3c_review}. The same holds for the tight-binding Hamiltonians (DFTB3-D3~\cite{dftb,dftb2,3ob,dftb3d3}, GFN2-xTB~\cite{gfnxtb,gfn2xtb}), especially GFN2-xTB  seems to profit a lot from error compensations.
Clearly, many more exchange correlation functionals exist, but benchmarking all variants exceeds the aim of this study.

Large scale dynamics studies of water at graphitic interfaces have been performed in the past years using classical water force fields combined with Lennard-Jones parameters for the oxygen-carbon interaction~\cite{Hummer:2001gd,Holt:2006kr,md_w@g_wetting,md_gwg,md_w@g_droplet,md_w@capillaries,md_w@g_angle}. Parameters have been adjusted to reference data from quantum chemistry on graphene-like clusters~\cite{LJ-water-carbon_mp2} or by reproducing experimental data like the contact angle of a water droplet on graphite~\cite{LJ-water-carbon_angle,LJ-water-carbon_wetting}. 
Our study has two major implications apart from an increased scepticism regarding the older theoretical reference data. 
First, we find that a Morse potential is much better suited to describe the water molecule-graphene interaction compared to Lennard-Jones potentials (Fig.~\ref{fig:pes} and Appendix~\ref{app:angle}). 
Furthermore, a coarse-grained model describing the interaction only as a carbon-oxygen interaction cannot describe both the 0-leg and 2-leg adsorption motifs simultaneously, i.e.\ the hydrogen-carbon interactions are mandatory for a qualitatively correct description.
Our new data will be valuable for further potential refinements, though for faithful predictions additional non-equilibrium geometries and high water coverages are needed.

\section{Analyzing the nature of the water-graphene interaction}
\label{sec:analyze}
\begin{figure}[hbt]
\centering
\includegraphics[width=0.49\textwidth]{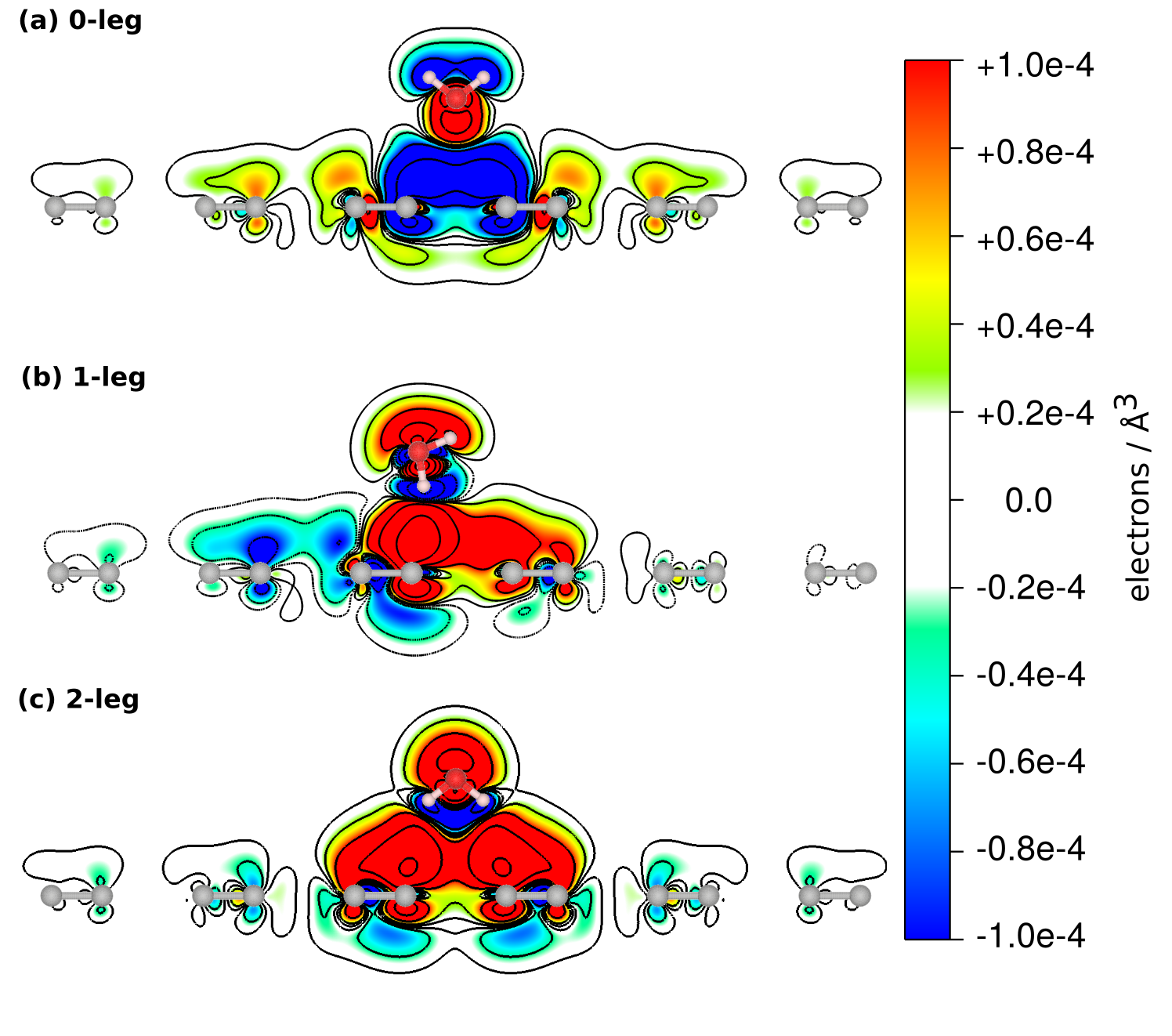}
\vspace{-0.3cm}
\caption{\label{fig:density} 
Electronic density change upon binding water on graphene in the equilibrium geometry for the (a) 0-leg, (b) 1-leg, and (c) 2-leg configurations.
Planes perpendicular to the surface that bisect the water molecules are shown. 
Red regions indicate density increase upon binding; blue regions indicate depletion.
Electronic density was calculated with DFT, using the PBE functional in a 5$\times$5 unit cell.
}
\end{figure}

We now briefly discuss the nature of the interaction between water and graphene.
As part of our analysis we examined how, at the DFT level, the electronic charge density rearranges upon 
creation of the adsorption bond. 
This is shown in  Fig.~\ref{fig:density} for the three binding motifs.
Charge density rearrangement plots such as this provide a pictorial representation of how the electron density rearranges upon adsorption. 
The key features revealed by Fig.~\ref{fig:density} are 
that: 
(i) The most extensive areas of charge rearrangement are on the water molecule and in the immediate vicinity of each adsorption site; 
(ii) The extent of the charge density rearrangement is fairly long ranged; carbon atoms as far as 8 \AA\ from the molecule exhibit some (albeit small) change in their charge density;  
and (iii) In all configurations the electronegative oxygen atom gains charge density, while the hydrogens lose charge density. 
The key difference between the 0-leg motif and the others is that in the 0-leg structure the negative oxygen ion points towards the graphene layer, and induces a charge loss in the local area surrounding it. 
Part of the charge is transferred to the water molecule,  while the rest is redistributed within the graphene layer. 
This charge rearrangement in the substrate is slightly more extensive in the 0-leg structure than in the two other binding motifs, which 
explains in part the strong variation of the 0-leg adsorption energy upon going from benzene to graphene.
Note that the qualitatively different nature of the adsorption bond for the 0-leg motif compared to the others
has an impact on the surface dipole moment and thus the
work function of the subtsrate. 
Indeed at the water coverage considered, the DFT PBE based work functions are 3.7, 4.5, and 4.8\,meV for the 0-leg, 1-leg and 2-leg motifs, respectively. 
The differences are quite pronounced and could in principle be observable experimentally. 
In addition, as the molecule rotates on or diffuses across the surface the nature of the charge rearrangement is likely to vary rapidly and so too the work function. 

A more quantitative analysis of the adsorption bond has been performed with symmetry adapted perturbation theory for the small substrate models (SAPT~\cite{pipisapt,jansensapt2,sapt_wire,sapt_wire2}). 
The results of this analysis are reported in the Supporting Information. 
The key conclusion from this analysis is that the electrostatic component to the interaction is attractive for the 1-leg and 2-leg motifs, while it is repulsive for the 0-leg motif on benzene. 
However, this repulsive interaction decreases significantly for larger substrate sizes and at the same time the attraction from vdW interactions increases. 
We can overall conclude from this analysis that the 1-leg and 2-leg motifs behave in a rather similar manner,  while the 0-leg motif has a quite distinct electrostatic interaction along with a more pronounced charge reorganization in the substrate.


\section{Conclusions}
\label{sec:conclusion}
To conclude, we have computed the adsorption of a water monomer on a periodic graphene sheet with DMC, CCSD(T), and RPA. 
By comparison to RPA and to L-CCSD(T) on smaller aromatic substrates, we have been able to resolve discrepancies between previously published adsorption energies. 
We have shown that different water orientations have quite distinct interactions with the graphene layer, but that ultimately they yield very similar binding energies. 
Cheaper computational methods such as DFT have shown that the potential energy surface for water on graphene is very smooth. 
However, here we show for the first time from accurate many-body electronic structure methods that there is almost no orientational dependence in the water monomer adsorption energy. 
Whilst more work is needed, this could have implications for and could partly explain the very low friction coefficient of liquid water on sp$^2$-bonded carbon. 
Our study also shows that DFAs employing atom-pairwise vdW corrections do a poor job at describing the delicate nature of this adsorption system. 
However, modern functionals like PBE-D3, PBE-D4, PBE-MBD, and rev-vdW-DF2 perform reasonably well with deviations to the DMC references below chemical accuracy. 
Even slightly higher accuracy can be achieved with the hybrid functional PBE0-D4. 
We hope that the benchmark provided here will be of value in larger scale \emph{ab initio} MD and classical MD studies of water on graphene and other sp$^2$-bonded carbon materials.
Together with previous studies using a range of many-body methods to study adsorption on hexagonal boron-nitride (hBN)~\cite{water_at_hbn}, carbon nanotubes~\cite{water@cnt_qmc}, clays~\cite{Zen:2016cc}, and lithium hydride~\cite{lih_surface,water@lih_qmc}, our study demonstrates that adsorption energies on extended surfaces can now be computed with sub-chemical accuracy. 
These electronic structure approaches are becoming a robust and reliable tool and have the prospect of been applied routinely to surface adsorption problems.
%


\begin{acknowledgments}
We are grateful for discussions with Francesco Paesani and Narbe Mardirossian. We thank Stefan Grimme for providing access to the \emph{dftd4} code.
J.G.B acknowledges support by the Alexander von Humboldt foundation within the Feodor-Lynen program.
A.Z. and A.M. are supported by the European
Research Council (ERC) under the European Union’s Seventh Framework
Program (FP/2007-2013)/ERC Grant Agreement 616121 (HeteroIce project).
A.Z. and A.M.’s work was also sponsored by the Air Force Office of Scientific Research, 
Air Force Material Command, US Air Force, under Grant
FA8655-12-1-2099. 
We are also grateful, for computational resources, to ARCHER UK
National Supercomputing Service, United Kingdom Car–Parrinello (UKCP)
consortium (EP/F036884/1), the London Centre for Nanotechnology, 
University College London (UCL) Research Computing, Oak Ridge Leadership Computing Facility (DE-AC05-00OR22725), 
and IT4Innovations Center of Excellence (CZ.1.05/1.1.00/02.0070 and LM2015070),
and the UK Materials and Molecular Modelling Hub, which is partially funded by EPSRC (EP/P020194/1).
A.G. and T.T. acknowledge funding from the European Research Council
(ERC) under the European Union’s Horizon 2020 research and innovation
program (grant agreement No 715594).  p-CCSD(T) calculations were
conducted on the HPC system COBRA of the Max Planck Computing and Data
Facility (MPCDF).
\\

\noindent
J.G.B and A.Z. contributed equally to this work.
\end{acknowledgments}


\appendix

\section{Contact angle from molecular dynamics simulations}
\label{app:angle}
\begin{figure}[hbt]
\centering
\includegraphics[width=0.4\textwidth]{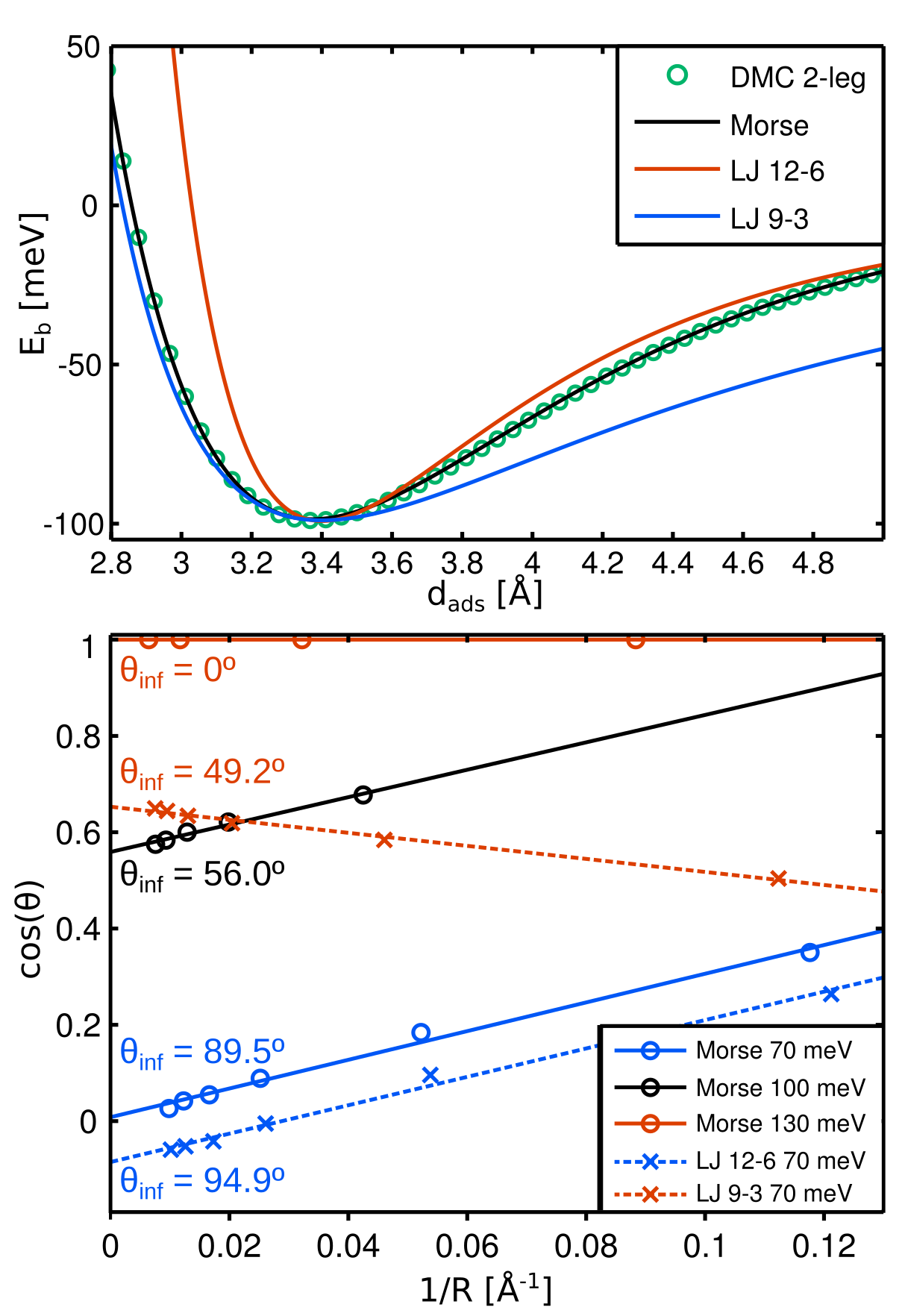}
\caption{\label{fig:angle_sim} 
Water adsorption modeled by a coarse-gained water model (mW) and different wall potentials. The upper panel shows the water-wall interaction potentials for an interaction strength of 100~meV. The lower panel shows the results for contact angles of different sized droplets, from which we extract the contact angle $\theta_\mathrm{inf}$ for an infinitely large droplet.
}
\end{figure}

The simulations to demonstrate the influence of the adsorption energy on the contact angle were done using the coarse-grained mW model~\cite{molinero_water_2009} of water and a Morse wall potential that acts as a function of the $z$ coordinate. The distance parameter was fitted to reproduce the DMC interaction curve for the 2-leg conformation. The resulting interaction curves, together with Lennard-Jones 9-3 and 12-6 wall variants, can be seen in figure \ref{fig:angle_sim}a.

To obtain contact angles we performed computations with the LAMMPS~\cite{plimpton_fast_1995} software, integrating the equations of motion with a timestep of 10~fs in the $NVT$ ensemble utilizing a 10-fold Nos\'{e}-Hoover chain~\cite{martyna_nosehoover_1992} with a relaxation time of 1~ps to realize a temperature of 300~K for a total time of 20 ns (an additional initial 10 ns to relax the droplet shape were discarded). From the trajectories we obtained the radial density profile of the liquid and fitted this to a spherical cap shape which yields the contact angle (we did not find any significant deviations from spherical shape). We performed this calculation for droplet sizes ranging from $\sim$ 600 to $\sim$ 70,000 molecules, fitting the size-dependent contact angles to the line-tension modified Young's equation~\cite{boruvka1977generalization}:
\begin{equation}
    \cos\left(\theta_R\right) = \cos\left(\theta_\mathrm{inf}\right) - \frac{1}{\gamma_\mathrm{lv}} \frac{\tau}{R} \nonumber
\end{equation}
which results in the contact angle for an infinitely large water droplet. Here, $\theta_R$ is the measured contact angle for a given droplet, $\theta_\mathrm{inf}$ is the contact angle of the infinite droplet, $\gamma_\mathrm{lv}$ is the liquid vapor surface tension (which does not need to be known for a fit), $\tau$ is the line tension and $R$ is the average radius of the contact area between the droplet and the wall. The results can be seen in Fig.~\ref{fig:angle_sim}b and show a stark influence of the contact angle to the used interaction strength and potential type. The simulation cells were periodic in $x$ and $y$ dimensions and with roughly 40~nm $\times$ 40~nm large enough to avoid self-interaction of the water molecules even for the complete wetting geometry in all cases but for the largest droplets.

\section{Computational Details}
\label{sec:appendix-compdetail}
\subsection{System}
\label{subsec:system}
We consider water adsorption in three different motifs, dubbed 0-leg, 1-leg, and 2-leg as defined in Fig.~\ref{fig:dmc_dario}. In order to be consistent with previous DMC calculations, the geometries are taken from PBE~\cite{pbe} optimizations, yielding the bond lengths $R(\text{C-C})=1.423$\,\AA\ within the graphene sheet, $R(\text{O-H})=0.972$\,\AA\ for the hydrogen atom pointing towards the surface in the 1-leg motif and $R(\text{O-H})=0.97$\,\AA\ otherwise. The hydrogen atoms of the water molecule have the usual bond angle of $\angle \text{HOH}=104.4^{\circ}$. The adsorption distance $d_{\text{ad}}$ is defined by the distance of oxygen to the graphene plane.  Molecular clusters  have been cut out of the periodic system and saturated with hydrogens. Here, we use fixed experimental bond lengths of 
$R(\text{C-C})=1.42$\,\AA,
$R(\text{O-H})=0.957$\,\AA,
and
$R(\text{C-H})=1.089$\,\AA.
Based on PBE-D3 calculations, we have tested that the different bond lengths give rise to a binding energy difference below 1.5\,meV.

\subsection{Density functional theory}
\label{subsec:dft}
DFT calculations have been performed with VASP\,5.4.4~\cite{vasp2,vasp3},
Turbomole\,7.2~\cite{turbomole_wire}, and
Orca\,4.0.1~\cite{orca}.
The numerical integration grids m4 and grid5
for Turbomole and Orca, respectively, have been used.
When applicable, the RI-J approximation was used with appropriate
auxiliary basis sets~\cite{riold1,riold3,ridft}.
Turbomole and Orca employ Gaussian type orbitals and
we have computed tightly converged interaction energies for the adsorption on benzene and coronene 
using PBE in a Dunning type aug-cc-pV5Z basis set expansion~\cite{dunning,augdunning}.
We confirmed that the basis set of def2-QZVPPD quality~\cite{qzvp,qzvp-thakkar} is indeed converged for DFT calculations with deviations 
from the larger basis set result below 2\,meV.

For the periodic system, a 5$\times$5 supercell is constructed with 20\,\AA\ vacuum in $z$-direction, energies are evaluated at the $\Gamma$-point, and a dipole correction is used in the $z$-direction.
VASP uses projector-augmented plane-waves (PAWs~\cite{paw1,paw2})
with an energy cutoff of 500\,eV. 
Neglecting the dipole correction leads to small deviations of about 2\,meV.
The convergence of the Brillouin zone sampling has been tested with a 2$\times$2$\times$1 $k$-grid  
and of the water coverage with a $\Gamma$-point calculation on a 10$\times$10 supercell (both corresponding to 100 
points in the Brillouin zone of the primitive graphene cell).
The difference between the converged supercell and the 5$\times$5 cell is with 5\,meV highest for the 1-leg motif,
which we mainly attribute to the remaining dipole interaction in the $x$-$y$-plane that is not present for the other adsorptions due to symmetry.
Consistent with RPA and DMC calculations, we use for the water reference the 2-leg geometry and thus accept a bias of up to 5\,meV.
For the finite sized systems, the dipole correction in the adsorption direction and 20\,\AA\ vacuum in all non-periodic directions is used.
This setting is compared to the PBE/aug-cc-pV5Z results for the molecular clusters and all interaction energies agree within 2\,meV.

For correcting missing long-range van der Waals interactions, we used
the D2~\cite{dftd2}, D3~\cite{dftd3}, D4~\cite{dftd3.5,dftd4}, VV10~\cite{vv10}, dDsC~\cite{ddsc}, TS~\cite{ts}, and MBD~\cite{ts-mbd}
schemes.
If not noted otherwise, D3 is used in its rational (Becke-Johnson) damping variant~\cite{dftd3bj,bjdamp} and includes the Axilrod-Teller-Muto type three-body term~\cite{atm1,atm2}.
As alternative approaches, we use the non-local density based functionals
of the vdW-DF1~\cite{vdwdf}, vdW-DF2~\cite{vdwdf2}, rev-vdW-DF2\cite{revvdwdf2}, and optB86b-vdW~\cite{optpbevdw} types.

\subsection{Random phase approximation}
\label{subsec:rpa}
We use RPA as implemented in Turbomole\,7.2~\cite{turbomole_wire} for molecular complexes
and a developer version of VASP\,6~\cite{vasp2,vasp3} for graphene.
The CP corrected complexation energy of $A$ and $B$ in basis sets $a$ and $b$ are computed via 
\begin{align}
 E^{\text{int}}_{CP} = E^{\text{int}} + E_{CP} \\
 E^{\text{int}}      = E(AB_{ab}) -E(A_{a}) - E(B_{b})\\
 E_{CP} =E(A_{a})-E(A_{ab}) + E(B_{b}) -E(B_{ab})
\end{align}
The Hartree and exact exchange energies (HXX@PBE) and the RPA correlation energy (RPA@PBE) are extrapolated to the 
basis set limit with optimized exponents~\cite{extrapolation}.
The final interaction energy is given by the extrapolated CP corrected energies $E^{\text{int}}_{CP} $ with a basis set error estimated as $|E_{CP}|/2$.
For the periodic system, PAWs with energy cutoff of 430\,eV are used. The results were extrapolated to the basis set limit assuming that errors drop off like one over the basis set size~\cite{RPA:Kresse2008,rpa_limit}. A quadrature with 8 grid points was used for the evaluation of the imaginary time and frequency integrations~\cite{vasp_rpaint2}.
The adsorption curves have been computed with 14\,\AA\ vacuum corrected with an increased vacuum of 20\,\AA\ at the equilibrium geometry.  Convergence of the first Brillouin zone sampling has been tested with additional calculations using 2$\times$2$\times$1 $k$-points and a 4$\times$4$\times$1 supercell  (the two settings correspond to 25 and 64 points in the Brillouin zone of a primitive graphene cell).

\subsection{Coupled cluster theory}
\label{subsec:cc}
\begin{figure}[hbt]
\centering
\includegraphics[width=0.5\textwidth]{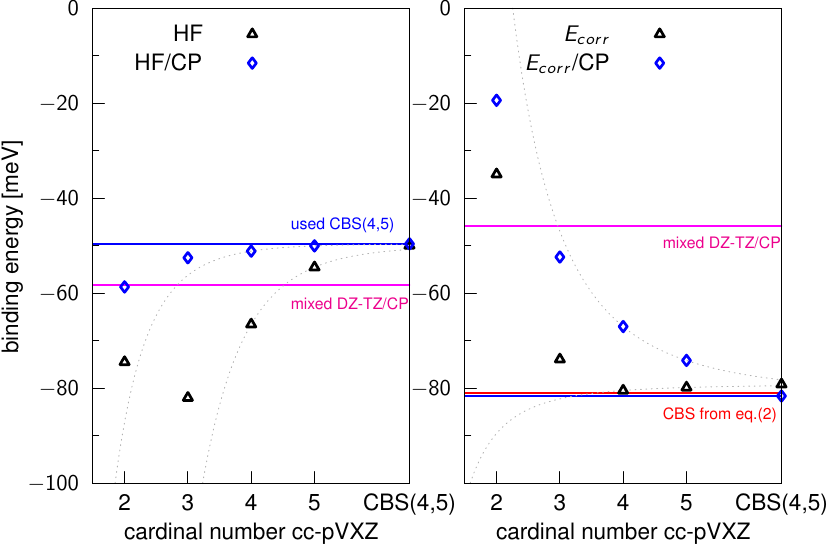}
\caption{\label{fig:basis} 
Binding energy separated into HF and correlation ($E^{\text{L-CCSD(T)}}_{\text{corr}}$) contributions for water adsorption on benzene in 2-leg geometry
($d_{\text{ad}}=3.51\,$\AA) and increasingly large basis set expansion.
Results from a mixed DZ/TZ basis set as in Ref.~\cite{water@graphene_paulus} and extrapolation according to eq.~\ref{eq:basis} are shown.
}
\end{figure}

We use the linear scaling domain based pair natural orbital CCSD(T) [denoted here as the LCCSD(T)] method~\cite{dlpnoccsdt} as implemented in the ORCA program package~\cite{orca}. The implementation has been optimized to use compact representations of all amplitudes and imposing block sparsity of tensors~\cite{dlpno_compact}.
Non-augmented basis sets are used in the CCSD(T) calculations to ensure numerically stable convergence on larger substrates. Though convergence of correlation energies with the employed basis sets is well studied~\cite{extrapolation_old,extrapolation_old2,s22},
it is mandatory to carefully test convergence for our target system and numerical settings.
In Fig.~\ref{fig:basis}, we show the convergence of uncorrected and CP-corrected Hartree-Fock (HF) and correlation $E_{\text{corr}}$ binding energies.
As usual for self-consistent field solutions, the plain HF interaction energies overestimate the binding due to basis set superposition error (BSSE).
The BSSE is quite effectively removed by the CP correction and for basis sets of about TZ quality can yield reasonably accurate results.
Still, the HF calculation is not the bottleneck in our study and we thus use direct extrapolations with QZ and 5Z basis sets~\cite{extrapolation} 
to minimize the associated errors.
The basis set artefacts on the correlation energy are more complex. Again, BSSE would lead to an overestimated binding energy,
while basis set incompleteness errors typically lead to an underestimated binding energy (missing correlation effects).
This can lead to uncorrected correlation energies that are closer to the basis set limit compared to the CP-corrected ones.
However, this trend is not clear and the convergence is much smoother using the CP-corrected energies (see Fig.~\ref{fig:basis}), which makes
the extrapolation more reliable.
In our study, the CCSD(T) correlation energies are extrapolated using the largest basis set results from L-CCSD(T)
and extrapolating it with RPA correlation energies in the multiplicative scheme
\begin{align}
\footnotesize
E_{\text{corr}}[\text{CC/CSB}] = E_{\text{corr}}(\text{CC/QZ}) \times \frac{E_{\text{corr}}(\text{RPA/CBS})}{E_{\text{corr}}(\text{RPA/QZ})}
\label{eq:basis}
\normalsize
\end{align}
The CP corrected energies are reported as final results; the non-CP corrected ones give an indication of the basis set completeness.
The extrapolation scheme has been chosen to minimize this error estimate (compared to e.g.\ the additive scheme).
For benzene adsorption, we compared this to direct cc-pV(QZ,5Z) extrapolations and deviations in binding energies are below 1\,meV 
(see Fig.~\ref{fig:basis}).
In contrast, a previously used combination of DZ and TZ basis sets (setting identical to Ref.~\cite{water@graphene_paulus}) have errors of about 30\,meV for the correlation energy. This partially cancels with errors on the HF energy, however, due to the functionally different convergence of HF and $E_{\text{corr}}$ we shouldn't expect this cancellation to be consistent for different system sizes.
Errors in our L-CCSD(T) energies due to pair-thresholds and non-canonical triples (T0) are estimated for benzene adsorption by comparison to conventional CCSD(T) in cc-pVTZ basis to be below 5\,meV, which is comparable to previous tests~\cite{water_at_hbn,dlpno_canonical_t}. Overall, the errors introduced by the truncation, the non-canonical triples, and the basis set extrapolation seem to be under control and below 10\,meV.

Periodic CCSD(T) calculations have been carried out following the strategy outlined in Ref.~\cite{pbc-ccsdt-Nlimit} employing a slightly smaller 4$\times$4 graphene cell. Periodic HF orbitals have been computed in a PAW basis with an kinetic energy cutoff for the plane wave basis of 500~eV, whereas virtual orbitals in the CCSD calculations are projected to a pseudized aug-cc-pVTZ basis set, in a PAW representation~\cite{booth2016}.
Perturbative triples (T) are evaluated using the smaller cc-pVDZ basis set to represent the virtual orbitals.
The CP corrected interaction energy is defined as
\begin{equation}
  E_{\text{int}} = E_{\text{H$_{2}$O+Graphene}} - E_{\text{H$_{2}$O}} - E_{\text{Graphene}}.
\end{equation}
Finite-size corrections have been computed at the CCSD level~\cite{pbc-ccsdt-fse} (-17, -16 and -19 meV for the 0-,1- and 2-leg structure, respectively). Finite coverage effects are corrected for at the HF level only using a $5\times5$ graphene cell (-2, 1 and 2 meV for the 0-,1- and 2-leg structure, respectively). Corrections to the vacuum size are computed using a supercell with a 30~\AA~vacuum distance at the MP2 level (6, 8 and 7 meV for the 0-,1- and 2-leg structure, respectively). A basis set correction is also included and is defined as the difference between the full plane-wave basis set calculation and the aug-cc-pVTZ one at the MP2 level (-4, -5 and -5 meV for the 0-,1- and 2-leg structure, respectively).

\subsection{Quantum Monte Carlo}
\label{subsec:qmc}
\begin{figure}[hbt]
\centering
\includegraphics[width=0.48\textwidth]{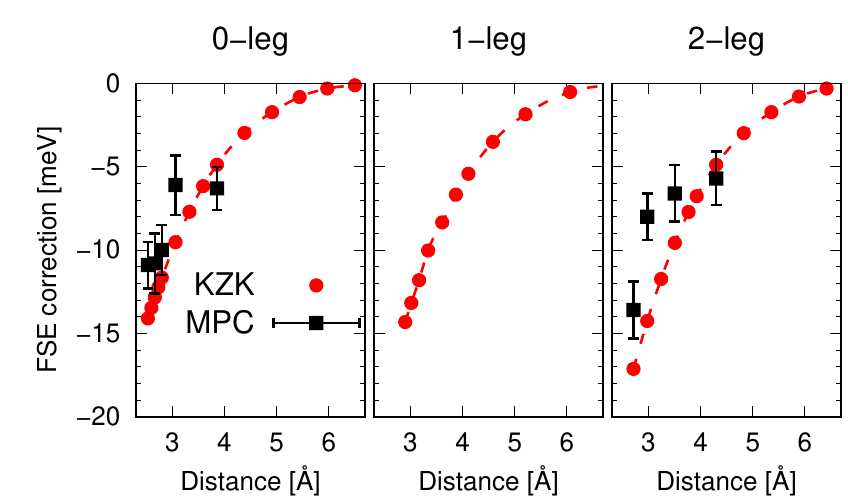}
\caption{\label{fig:DMCfse} 
FSE correction in DMC for the 
interaction energy of water on graphene, as obtained using KZK~\cite{KZK:prl2008} and MPC~\cite{MPC:Fraser1996,MPC:Will1997,MPC:Kent1999} approaches, shown as a function of the distance between the oxygen of water and the graphene sheet.
}
\end{figure}

DMC calculations were performed with the {\sc CASINO} code~\cite{casino}. 
The adsorption energy $E_\text{ad}$ 
is calculated 
as prescribed in Eq.~\ref{eq:ead},
with $d_\text{far} \sim 10$~\AA{}.
This evaluation is more efficient than the use of the separate fragments in place of the far away configuration, as it reduces the time step bias \cite{sizeconsDMC}. However, the system at $d_\text{far}\sim 10$~\AA{} could have a little residual interaction energy, which was evaluated via L-CCSD(T) to be 2.2~meV for the water-benzene system and 5.5~meV for the water-coronene system. Thus, the DMC evaluations were corrected for this interaction energy, in order to facilitate the comparison with the binding energies obtained via L-CCSD(T) and RPA, which took as reference the energy of the isolated fragments. For the graphene adsorption this residual interaction diminishes well below 1\,meV as estimated by DFT calculations (PBE-D3 and PBE-MBD).
Similar to Refs.~\cite{water@benzene_dmc, water@graphene_dmc}, we used Dirac-Fock pseudopotentials~\cite{trail05_NCHF, trail05_SRHF} with the locality approximation~\cite{mitas91}. 
Single particle wave functions were obtained using DFT with the LDA functional and a plane-wave cutoff of 600 Ry,  re-expanded in terms of B-splines~\cite{alfe04} with the natural grid spacing $a=\pi/G_{\rm max}$, where $G_{\rm max}$ is the magnitude of the largest plane wave in the expansion.  
The Jastrow factor used in the trial wavefunction of the system included a two-body electron-electron (e-e) term; two-body electron-nucleus (e-n) terms and three-body electron-electron-nucleus (e-e-n) terms specific for any atom type. 
The variational parameters of the Jastrow have been optimized in order to minimize the variational variance.
The time step dependence has been investigated explicitly considering values of $\tau$ ranging from $10^{-1}$~au to $10^{-3}$~au for a subset of configurations, as reported in the Supporting Information. 
Production calculations for water-benzene and water-coronene systems used a time step of 0.01~au,
and in water-graphene we used $\tau=0.025$~au.
These values give a bias smaller than the stochastic error. 
DMC calculations were performed with a population of tens of thousands of walkers or more.
We tested the population bias, which appears to be negligible with respect to the stochastic error in the production calculations (the population bias becomes of the order of a few meV only for a population of a few hundred walkers) as shown in the Supporting Information. %
In this work we are evaluating the 
interaction  energy as the difference of two configurations  both affected by FSE, thus we will benefit from a large FSE cancellation, as observed in other systems~\cite{Zen:2016cc, water_at_hbn}.
Similar to Ref.~\onlinecite{Zen:2016cc}, we have estimated the residual FSE correction using the approach of Kwee, Zhang and Krakauer~\cite{KZK:prl2008} (KZK). 
In a subset of the configurations we checked the reliability of KZK against the more accurate (and computationally more expensive) model periodic Coulomb (MPC) approach~\cite{MPC:Fraser1996,MPC:Will1997,MPC:Kent1999}.
The estimated FSE correction on DMC binding values are reported in fig.~\ref{fig:DMCfse}.

\section{Fit with Morse potential}
\label{sec:appendix-morses}
\begin{figure}[htb]
\centering
\includegraphics[width=0.48\textwidth]{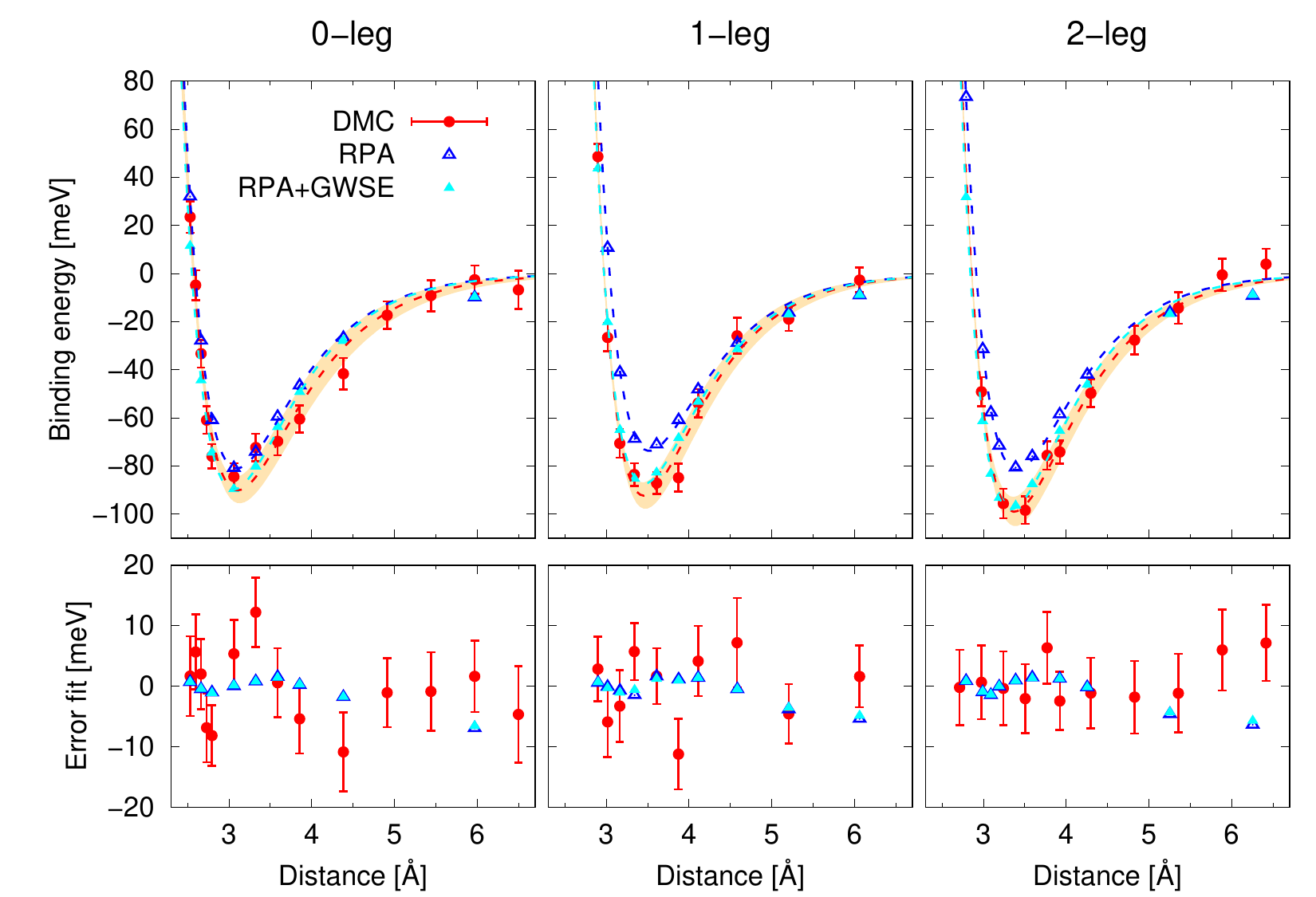}
\caption{\label{fig:morseWG} 
Adsorption energy (top) of water on graphene  and difference (bottom) with respect to the fitted Morse potential (dashed lines).
}
\end{figure}
\begin{figure}[tbh]
\centering
\includegraphics[width=0.48\textwidth]{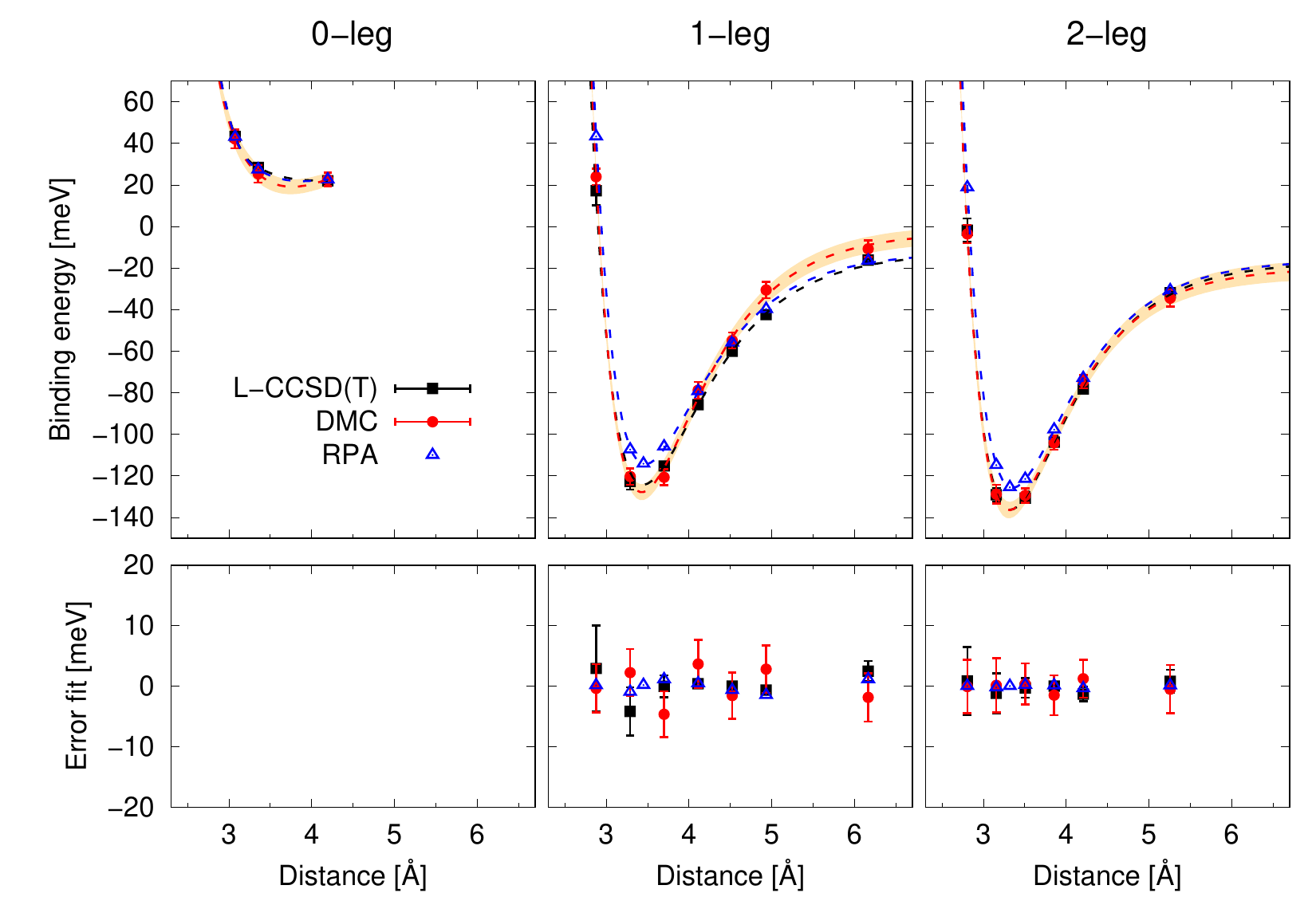}
\caption{\label{fig:morseWBZ} 
Adsorption energy (top)  of water on benzene  and difference (bottom) with respect to the fitted Morse potential (dashed lines).
A cubic spline is used for the non-binding 0-leg motif.
}
\end{figure}
\begin{figure}[tbh]
\centering
\includegraphics[width=0.48\textwidth]{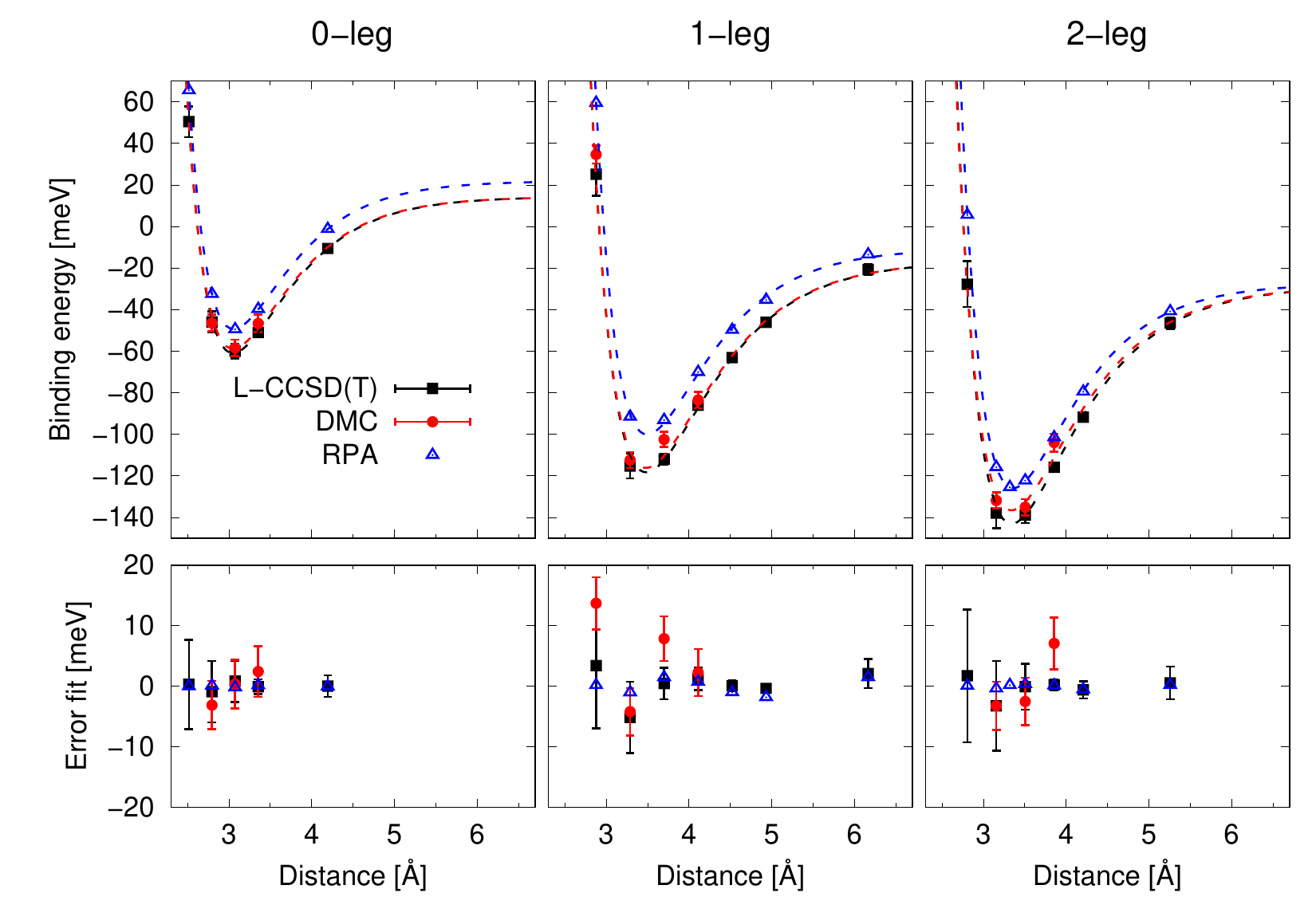}
\caption{\label{fig:morseWCORO} 
Adsorption energy (top)  of water on coronene  and difference (bottom) with respect to the fitted Morse potential (dashed lines).
}
\end{figure}

The dissociation curve of water on graphene was fitted using a 
Morse potential
\begin{align}
\footnotesize
E_{\text{b}}(d) = D_e \left[ \left(1 - e^{-a(d-d_{\text{min}})} \right)^2 - 1 \right] +D_{\text{o}} \,,
\label{eq:morse}
\normalsize
\end{align}
where $d$ is the distance of the water oxygen from graphene (see Fig~\ref{fig:dmc_dario}), and the parameters 
$D_e$, $a$, $d_{\text{min}}$ and $D_o$ are fitted independently for each of the three motifs.
Lower panels of Fig.~\ref{fig:morseWG} show that in the range of distances considered in this work the agreement is optimal for $D_{\text{o}}=0$,
as the RPA and RPA+GSWE values are within 2~meV from the fit for $d$ close to $d_\text{min}$.
The values of $E_{\text{ad}}$ and $d_\text{min}$ reported in Table~\ref{tab:energies}, and the associated errors, are obtained from the fit, and in particular from the parameters $-D_e+D_o$ and $d_{\text{min}}$.
A Morse potential is used also for the binding curve of benzene, see Fig.~\ref{fig:morseWBZ}, and coronene, see Fig.~\ref{fig:morseWCORO}.
However, in these two cases the Morse potential needs an offset $D_o$, as the binding curve converge to zero with a long tail.
The lower panels in Figs.~\ref{fig:morseWBZ} and \ref{fig:morseWCORO} show optimal agreement between the computed values and the Morse potential in the considered range of distances.
The values of $E_{\text{ad}}$ reported in Table~\ref{tab:energies} is calculated as $E_{\text{ad}}= -D_e+D_{\text{o}}$, and the associated error is the square root of the sum of squares of the fitting errors. This is a conservative estimation of the error, as $D_e$ and $D_{\text{o}}$ are anti-correlated.
All Morse parameters are reported in the supporting information, which we recommend for benchmarking approximate methods.
%


\begin{thebibliography}{136}
\bibitem [{sub()}]{substrate-radius}%
  \BibitemOpen
  \href@noop {} {}\bibinfo {note} {We define the substrate size as the circle
  in the substrate plane with minimal radius, where all atom positions are
  within the circle (see inset of Fig.~\ref{fig:convergence}).}\BibitemShut
  {Stop}%
\bibitem [{dmc()}]{dmc_error}%
  \BibitemOpen
  \href@noop {} {}\bibinfo {note} {Note that this trend is still possible
  within the stochastic error of the DMC data.}\BibitemShut {Stop}%
\bibitem [{dft()}]{dftd4}%
  \BibitemOpen
  \href@noop {} {}\bibinfo {note} {DFT-D4 model will be published elsewhere. It
  is based on Ref.~\cite{dftd3.5} and includes all dipole many-body
  orders.}\BibitemShut {Stop}%
\end{thebibliography}

%


\end{document}